\documentstyle[preprint,pre,aps,eqsecnum,epsf]{revtex}

\begin{document}
\draft


\title{Incorporating Radial Flow in the Lattice Gas Model for Nuclear 
Disassembly}

\author{C. B. Das and S. Das Gupta}
\address{
Physics Department, McGill University,
3600 University St., Montr{\'e}al, Qu{\'e}bec \\ Canada H3A 2T8\\ }

\date{ \today } 

\maketitle

\begin{abstract}
We consider extensions of the lattice gas model to incorporate radial
flow.  Experimental data are used to set the magnitude of radial flow.
This flow is then included in the Lattice Gas Model in a microcanonical
formalism.  For magnitudes of flow seen in experiments, the main 
effect of the flow on observables is a shift along the $E^*/A$ axis.

\end{abstract}



It is expected that when nuclei disintegrate after heavy ion collisions,
there will be a radial flow in the disintegrating system in addition
to chaotic motion which is usually described by thermal motion.  This
was first proposed for collisions in the Bevalac \cite{Siemens} but is
expected and seen in collisions at lower energies as well 
\cite {Hsi,Lisa,Williams}.  The amount of radial flow is larger for
central collisions.

In this paper we address the issue of incorporating
radial flow in statistical models for nuclear disassembly.  This is
automatically taken into account in models based on transport equations
such as BUU (Boltzmann-Uehling-Uhlenbeck)
\cite {Bertsch}.  But in many statistical models such
as the SMM (statistical multifragmentation model \cite {Bondorf}), 
thermodynamic model \cite {Dasgupta} or 
the microcanonical model \cite {Gross}
the flow can only be included {\it a posteriori}.  The idea here would be that
the energy which is lost in radial flow is lost for thermalisation, thus
essentially less energy is available for thermal disassembly.  While
this idea is certainly quite attractive, radial flow may do more than 
just take away energy.  As far as we know this was first pointed out
in \cite{Kunde}.
It is this aspect that we study quantitatively here.

Some additional insight can be gained by incorporating a radial flow 
in the Lattice Gas Model (LGM) which
is being applied more and more to fit experimental data \cite {Chomaz,Sam}.
In the usual formulation of the LGM, equilibrium statistical
mechanics is done before composites are calculated \cite {Pan}.  We 
combine statistical mechanics with radial flow
and then examine how it affects the
composite production.  The important issue here is the relative kinetic
energy of two nearest neighbours.  If this is less than the attractive
bond, the two nearest neighbours will be part of the same cluster.  Now,
if one has a radial flow which diverges outward from the centre, the
flow will affect the relative kinetic energy and hence the composite
production.  The argument of merely some energy being unavailable
for thermalisation is strictly valid when the collective velocity
is the same for every nucleon.

We are not claiming that radial flow arises in
LGM in a fundamental way.  But it can be included with a reasonable
prescription.  Inclusion of radial flow in a model similar to LGM 
was considered by Elattari et al \cite {Richert}.  Here we base
our calculations on experimental data.  The data used here is from a 
work by Williams et al. \cite{Williams}. In that paper experimental data
of average radial flow is plotted as a function of excitation energy
per nucleon ($E^*/A$).  This is converted here to $E_f/A$ against 
$E^*/A$ (see Fig. 1) where $E_f/A$ is the flow energy per nucleon.
Our calculation, by construct, will reproduce this curve.
The model we use is this.  From Fig. 1 we can also construct a 
$E_f/A$ against $E_{stat}/A$ where $E_{stat}/A=E^*/A-E_f/A$.  The
part $E_{stat}/A$ is generated by the LGM.  In LGM we generate events
which pertain to a $E_{stat}/A$.  If there were no flow then from
these events we would generate clusters and compare with experiments.
But since experiments dictate that there is also energy tied up
with flow we impose a flow energy on each event.  The amount of flow
energy is taken from the experimental $E_f/A$ against $E_{stat}/A$
curve.  We take the flow velocity to be proportional to the distance
from the center of mass of the exploding nucleus.  Since in an event
the position of each nucleon is known, this can be done uniquely.
This is the principle of this calculation; below we provide some more
details.

The most straightforward approach would be to use a microcanonical
Lattice gas model.  This is as simple as a canonical Lattice gas
model calculation (see ref. \cite{Das1}).
Our simulations are done for $A=84, N=48, Z=36$.  We use a $6^3$ lattice.
The neutron-proton bond is -5.33 MeV; like particle bonds are set at
0.  Coulomb interaction between protons is taken into account as in
\cite {Sam}. We use Metropolis Monte-Carlo simulations to obtain
microcanonical samplings.
We start from a suitable initial lattice configuration which gives an
interaction energy $E_{pot}$.  The statistical energy $E_{stat}$
is fixed.  The statistical kinetic energy for this configuration
is then $E_{stat}+E_{ground}-E_{pot}=E_{kin}$.  
The phase space $\Omega_k(E_{kin})$ available for
this kinetic energy is well-known:

$\Omega_k(E_{kin})=\int d^3p_1d^3p_2.....d^3p_A
\delta (E_{kin}-\sum \frac{p_i^2}{2m})=
\frac {(2\pi m)^{3A/2}}{\Gamma(3A/2)}(E_{kin})^{3A/2-1}$

We now try to switch to a different configuration in the lattice.  As a
result, the potential energy in this new configuration would change to a new
value $E_{pot}'$.  In this configuration, since we are doing a 
microcanonical simulation, the statistical kinetic energy would have to 
adapt to a new value:$ E_{kin}'=E_{stat}+E_{ground}-E_{pot}'$.  
Correspondingly, the new phase-space
will be $\Omega_k(E_{kin}')$.  If this is bigger than $\Omega_k(E_{kin})$,
the switch is made.  Otherwise the switch is made with a probability
$\frac{\Omega_k(E_{kin}')}{\Omega_k(E_{kin})}$.  After many such switches
(some successful and some not) we accept an event. 
We have to now assign momenta to the nucleons.  Let the total statistical
kinetic energy of the $N$ nucleons be $\tilde E_{kin}$ for the chosen
event.  This has to be shared between the nucleons based solely upon
phase-space.  This can be done following this procedure.  Choose a sphere
of radius $P$.  Do a Monte-Carlo sampling on $N$ nucleons for uniform
distribution in this sphere.  This means fixing $p, \theta_p$ and
$\phi_p$ for each particle from $p=P(x_1)^{1/3}, cos\theta_p=1-2x_2$
and $\phi_p=2\pi x_3$ where $x_1, x_2$ and $x_3$ are random numbers in
the domain 0 to 1.  Finally normalize $P$ so that the total statistical kinetic
energy equals $\tilde E_{kin}$.  If there were no flow, we would now
do cluster decomposition for this event to compare directly
with experiment.  The standard prescription
is that two nearest neighbor nucleons are part of the same cluster
if the kinetic energy of relative motion is unable to overcome the
attractive bond: $p_r^2/2\mu+\epsilon<0$.
To include flow,  we add
radial momenta to the statistical momenta already generated, thus
get new momenta and then apply the prescription above for composites.
The model for flow we
adopt is that nucleons will have, apart from thermal motion, a
momentum which is proportional to the distance from the centre of
mass of the cluster:  $\vec p_f(i)=c\times(\vec r(i)-
\vec r_{cm})$.  Here $\vec r(i)$ denotes the position of the nucleon
in the lattice and $\vec r_{cm}=\frac{\sum \vec r(i)}{A}$.
The constant $c$ is adjusted so that the total flow energy
adds up to the pre-assigned value that we choose from the
experimental $E_f$ {\it vs.} $E_{stat}$ curve.  It should be
pointed out that because we have many particles, the vector 
addition of flow momentum to the thermal momentum just leads
to a scalar addition of energy, i.e., after addition of flow
momenta the total energy remains, to a very good accuracy, 
$E_f+E_{stat}$.  Thus we automatically satisfy the experimental data.

We obtain clusters both in the model just described and in a standard
LGM microcanonical model without flow.  To compare we extract an
approximate exponent $\tau_Z$ ($Y(Z)\propto Z^{-\tau_Z}$) in both the
models and plot it against $E^*/A$ (Fig.2).  The main influence of the flow
is a shift along the $E^*/A$ axis but this shift is not constant.  
In the same figure we also plot $\tau_Z$ against $E_{stat}/A$.
If the only influence of the flow was to take out some energy but
otherwise leave the cluster production unchanged the two curves
in the lower part of Fig. 2 would be on top of each other.
The fact that the dashed curve trails the solid curve as a function
of $E_{stat}/A$ supports the conjecture made in \cite{Kunde}, this
time in a fully microscopic model.

For the same value of $\tau_Z$ the cluster decomposition is basically
the same (Fig.3)

In summary, we have incorporated the radial flow in nuclear
Lattice Gas model with a reasonable prescription.
For flow energy between 30 to 50$\%$ of thermal energy, as
suggested by experiments, the inclusion of flow does not affect the fragment 
production in a profound way in intermediate energy heavy-ion collisions.
While this calculation establishes the framework to include radial flow
in the lattice gas model, in future we hope to do calculations for specific
experimental cases including very large systems such as considered in
\cite{Kunde}.

This work was supported in part by the Natural Sciences and Engineering 
Council of Canada and by {\it le Fonds pour la Formation de chercheurs
et l'Aide \`a la Recherche du Qu\'ebec}.  We thank S. K. Samaddar for
use of his computer code which was re-adopted for this problem.

\newpage

\epsfxsize=4.0in
\epsfysize=6.0in
\par
\centerline{\epsffile{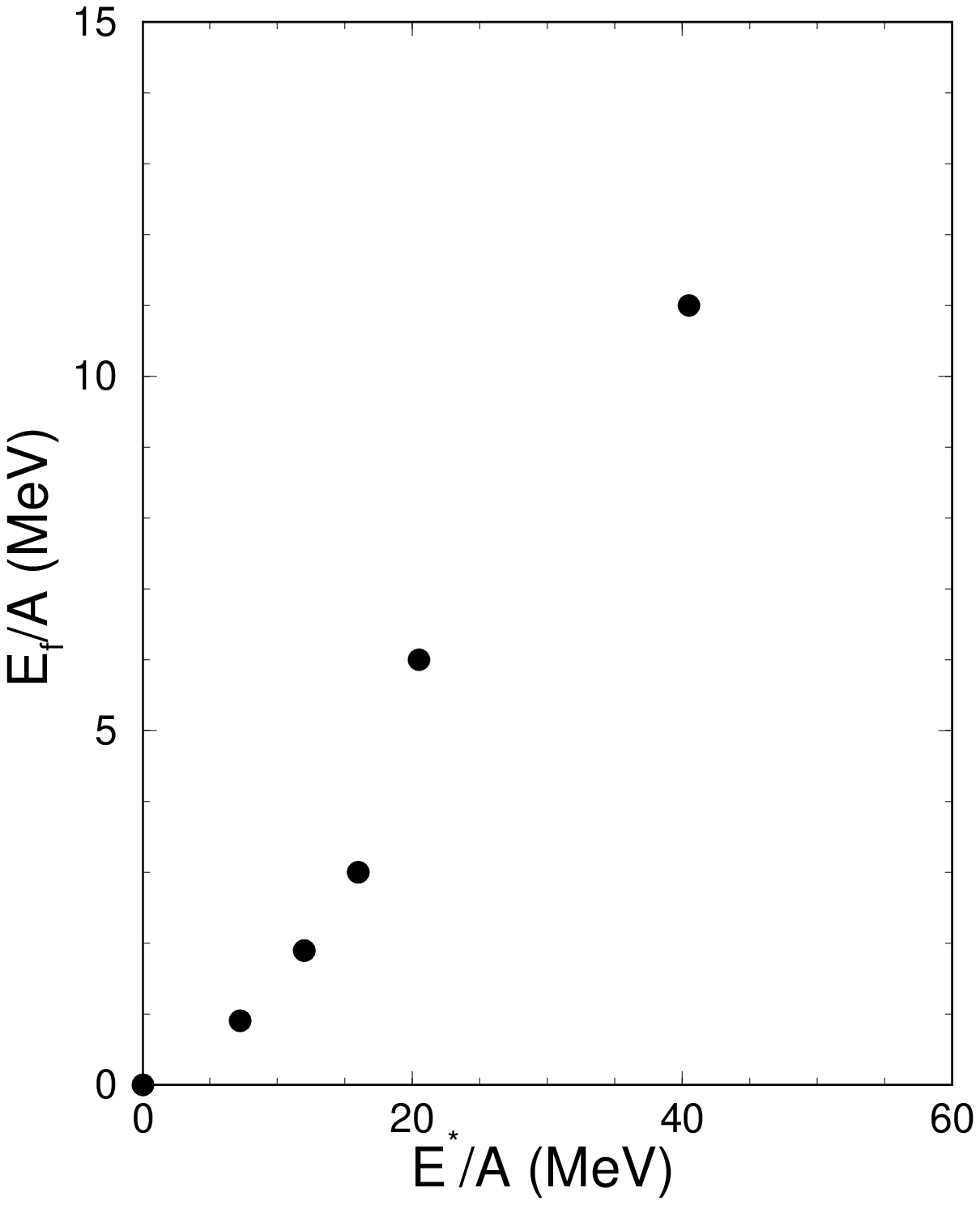}}
\par
\vspace{0.5in}
Fig.1: Experimental data on flow energy per nucleon, $<E_{f}/A>$  plotted 
as a function of excitation energy per nucleon, $E^*/A$ from \cite{Williams}.
The error bars are not shown.

\newpage

\epsfxsize=5.5in
\epsfysize=6.5in
\par
\centerline{\epsffile{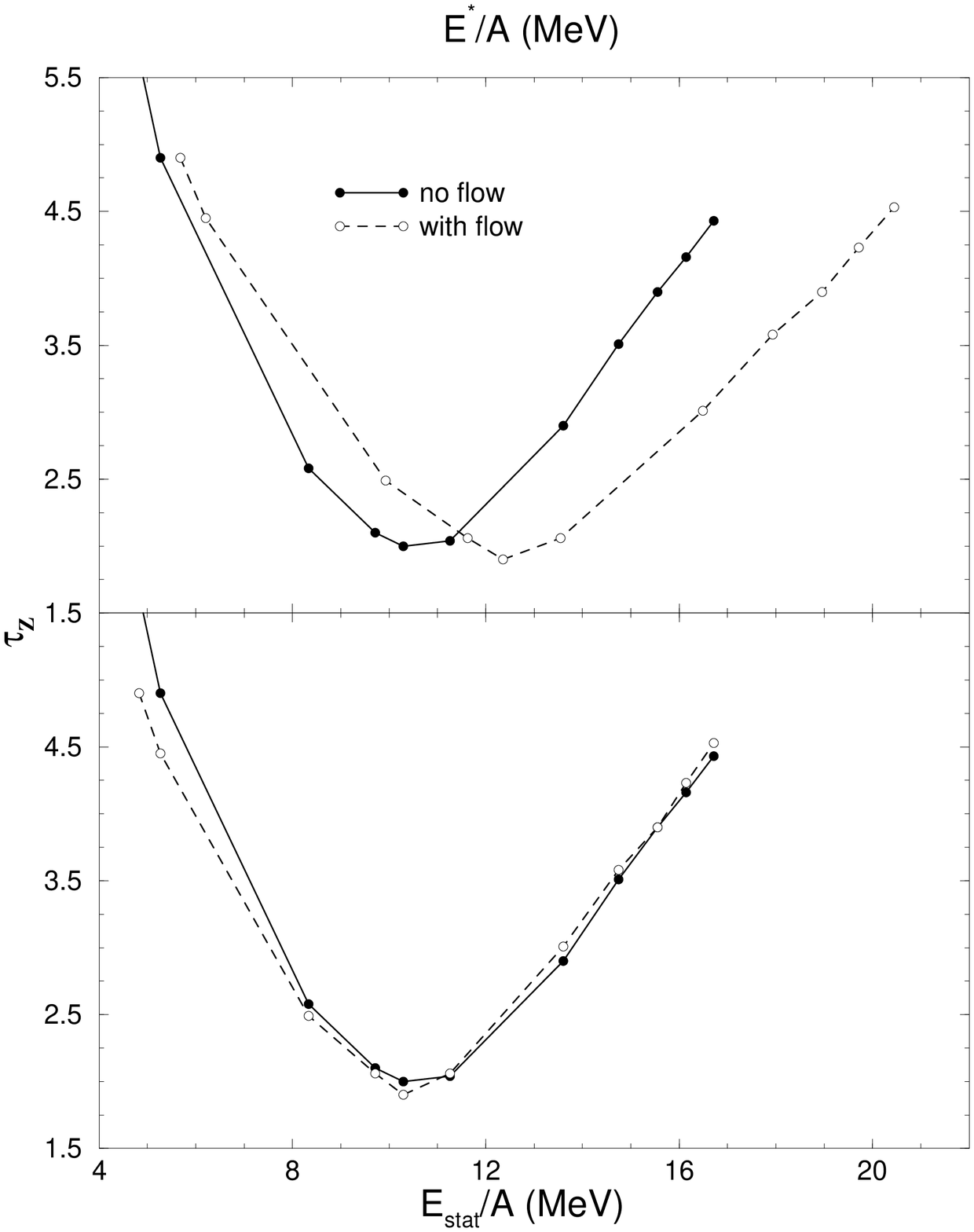}}
\par
\vspace{0.5in}
Fig. 2: The extracted values of the exponent $\tau_Z$, 
with and without flow, as a function of $E^*/A$ (upper panel).  In doing
the calculation with flow, we used interpolation between experimental
points (Fig.1) and obtained an analytical expression for $E_f/A$ against
$E^*/A-E_f/A$.  In the lower panel we plot $\tau_Z$ for both the models
but for fixed $E_{stat}/A$.  The fact that the two curves are not exactly 
on top of each other shows that flow does more than just takes away some 
energy.
\newpage

\epsfxsize=4.5in
\epsfysize=6.5in
\par
\centerline{\epsffile{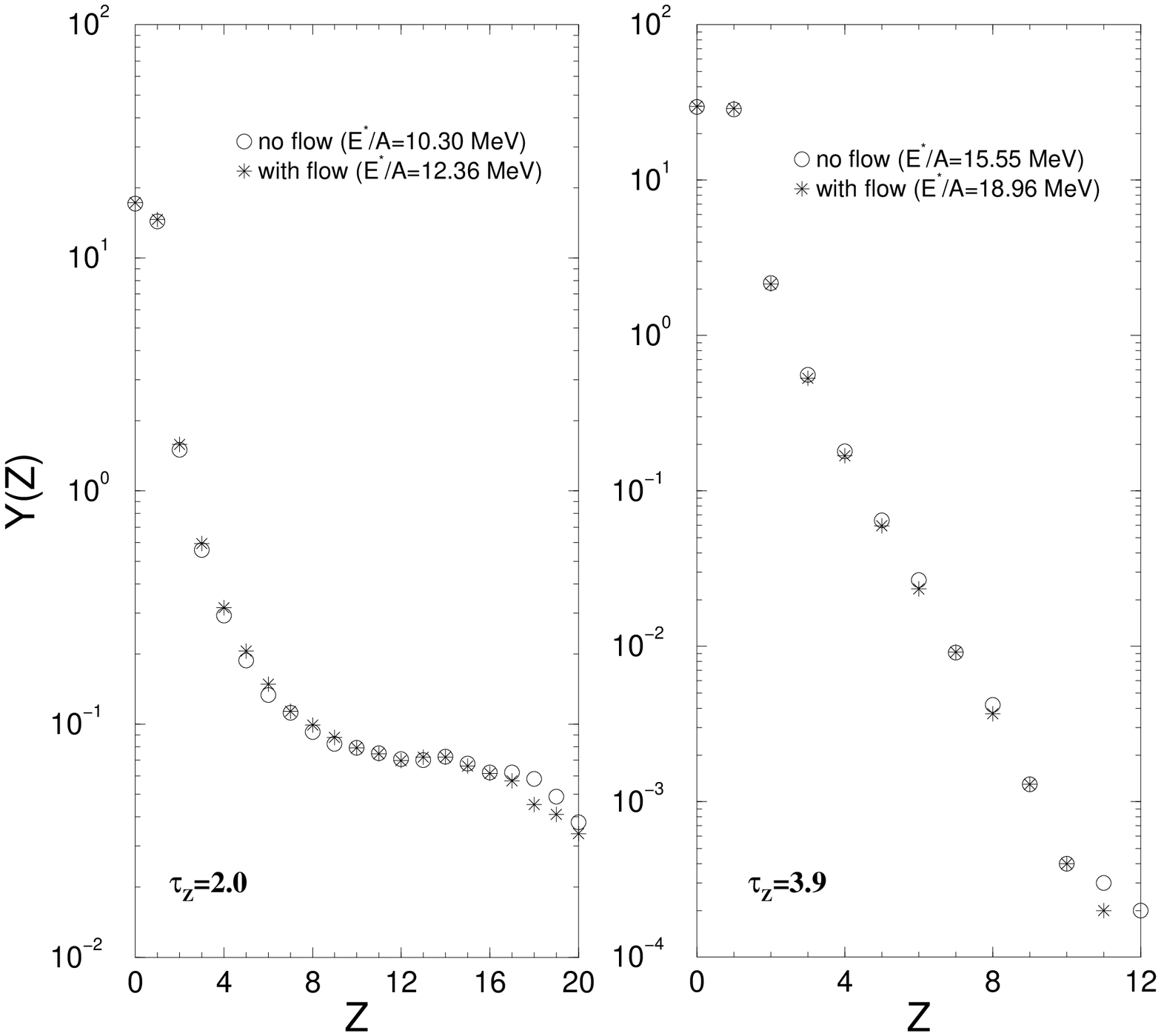}}
\par
\vspace{0.5in}
Fig. 3: Charge yields of the system $A=84$ with and without flow, at
different excitation energies which have same exponents.

\end{document}